# Vanadium oxide metal-insulator phase transition in different types of one-dimensional photonic microcavities

Francesco Scotognella
Dipartimento di Fisica, Politecnico di Milano, piazza Leonardo da Vinci 32, 20133 Milano, Italy
Email: francesco.scotognella@polimi.it

**Abstract**
The optical properties of vanadium dioxide ($VO_2$) can be tuned via metal-insulator transition. In this work different types of one-dimensional photonic structure-based microcavities that embed vanadium dioxide have been studied in the spectral range between 900 nm and 2000 nm. In particular, $VO_2$ has been sandwiched between: i) two photonic crystals made of $SiO_2$ and $ZrO_2$; ii) two aperiodic structures made of $SiO_2$ and $ZrO_2$ that follow the Thue-Morse sequence; iii) two disordered photonic structures, made of $SiO_2$ and $ZrO_2$ in which the disorder is introduced either by a random sequence of the two materials or by a random variation of the thicknesses of the layers; iv) two four material-based photonic crystals made of $SiO_2$, $Al_2O_3$, $Y_2O_3$, and $ZrO_2$. The ordered structures i and iv show, respectively, one and two intense transmission valleys with defect modes, while the aperiodic and disordered structures ii and iii show a manifold of transmission valleys due to their complex layered configurations. The metal-insulator transition of $VO_2$, controlled by temperature, results in a modulation of the optical properties of the microcavities.

**Introduction**
Crystalline vanadium dioxide ($VO_2$) shows a thermochromic phase transition around 68 °C (341 K) [1]. The phase transition is related to a structural crystal change from a monoclinic insulating phase to a tetragonal (rutile) metallic phase [2,3]. From an optical point of view, the phase transition of $VO_2$ results in a change from an insulating semi-transparent material from a metallic more lossy and reflective material [3,4]. $VO_2$ phase transition can be exploited for several applications, such as smart windows, steep-slope devices for micro-electronics, neuromorphic computing devices, and reconfigurable radiofrequency switches [5,6].
A method to utilize $VO_2$ switchable optical properties is the integration of such material is its integration in photonic crystals [7–9]. In photonic crystals, the periodic modulation of the refractive index in one, two or three dimensions gives rise to energy regions in which light is not transmitted through the crystal. The integration of materials with switchable optical properties in the infrared, such photochromic polymers [10] and infrared plasmonic nanomaterials [11,12], in one-dimensional photonic crystals has been proposed [13].
In this work, we propose different types of one-dimensional photonic microcavities [14], in which a layer of $VO_2$ is embedded between two photonic crystals, two aperiodic Thue-Morse photonic structures, two disordered photonic structures, and two four-material based photonic crystals. The wavelength dependent refractive indexes of all the employed materials have been used. The light transmission of the microcavities has been simulated via the transfer matrix method. The modulation of the transmission spectra of the microcavities due the $VO_2$ metal-insulator phase transition has been highlighted.

**Methods**
The light transmission of the different microcavities in the spectral range between 900 nm and 2000 nm has been studied with the transfer matrix method [15–17]. The system is glass/multilayer/air with light impinging the sample surface orthogonally. The characteristic matrix of the multilayer is written as

$$M = \begin{bmatrix} M_{11} & M_{12} \\ M_{21} & M_{22} \end{bmatrix} = \prod_{j=1}^{N} \begin{bmatrix} \cos\left(\frac{2\pi}{\lambda} n_k(\lambda) d_k\right) & -\frac{i}{n_k(\lambda)} \sin\left(\frac{2\pi}{\lambda} n_k(\lambda) d_k\right) \\ -i n_k(\lambda) \sin\left(\frac{2\pi}{\lambda} n_k(\lambda) d_k\right) & \cos\left(\frac{2\pi}{\lambda} n_k(\lambda) d_k\right) \end{bmatrix} \quad (1)$$

With $k=(1,...,N)$ and $N$ number of layers. $d_k$ and $n_k(\lambda)$ are the thickness and the wavelength dependent refractive index of the $k$th layer, respectively. The light transmission is written as

$$T = \frac{n_0}{n_g} \left| \frac{2 n_g}{(M_{11} + M_{12} n_0) n_g + (M_{21} + M_{22} n_0)} \right|^2 \quad (2)$$

With $n_g$ the refractive index of glass and $n_0$ the refractive index of air ($n_g = 1.46$; $n_0 \cong 1$). The light transmission has been calculated in the selected spectral range with steps of 0.25 nm. The wavelength dependent refractive index $n_k(\lambda)$ can be written with the Sellmeier equation

$$n_k^2(\lambda) - 1 = \sum_{j'=1} \frac{A_{j'} \lambda^2}{\lambda^2 - B_{j'}^2} \quad (3)$$

The parameters $A_{j'}$ and $B_{j'}$ are reported in Table 1.

| Material | A$_1$ | B$_1$ | A$_2$ | B$_2$ | A$_3$ | B$_3$ | Ref. |
|---|---|---|---|---|---|---|---|
| SiO$_2$ | 0.6961663 | 0.0684043 | 0.4079426 | 0.1162414 | 0.8974794 | 9.896161 | [18,19] |
| Al$_2$O$_3$ | 1.023798 | 0.0614482 | 1.058264 | 0.1106997 | 5.280792 | 17.92656 | [20] |
| Y$_2$O$_3$ | 2.578 | 0.1387 | 3.935 | 22.936 | - | - | [21] |
| ZrO$_2$ | 1.347091 | 0.062543 | 2.117788 | 0.166739 | 9.452943 | 24.32057 | [22] |

**Table 1**. Parameters of the Sellmeier equation for SiO$_2$, Al$_2$O$_3$, Y$_2$O$_3$, and ZrO$_2$.

The wavelength dependent refractive indexes of vanadium dioxide in the metal phase and in the insulating phase (30 °C) and in the metallic phase (100 °C) have been taken from Ref. [2].

**Results and Discussion**

In Figure 1 the light transmission spectrum for the microcavity (SiO$_2$/ZrO$_2$)$_5$/VO$_2$/(ZrO$_2$/SiO$_2$)$_5$ is shown. The black solid curve is related to VO$_2$ in the insulating phase, while the blue dashed curve is related to VO$_2$ in the metallic phase. The two phases correspond to a temperature of the material of 30 °C for the insulating phase and a temperature of 100 °C for the metallic phase, respectively, as reported in Ref. [2]. In the microcavity the thickness of the silicon dioxide layers is 220 nm, the thickness of the zirconium dioxide layers is 165 nm, while the thickness of the vanadium dioxide layers is 55 nm.

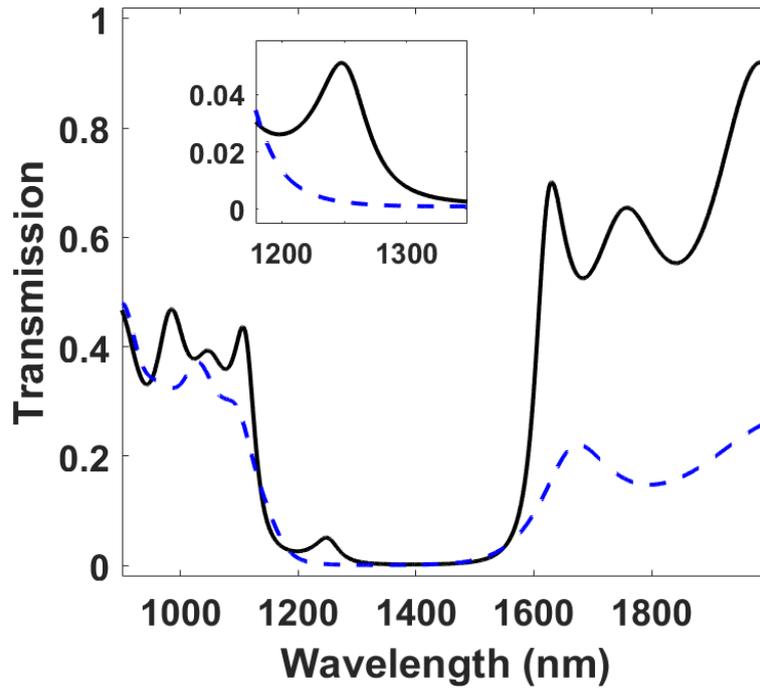

**Figure 1**. Light transmission spectra for the microcavity $(SiO_2/ZrO_2)_5/VO_2/(ZrO_2/SiO_2)_5$ with $VO_2$ in the insulating phase (at 30 °C, black solid curve) and in the metallic phase (at 100 °C, blue dashed curve).

In the microcavity configuration, the insulating $VO_2$ based microcavity shows a defect mode at around 1250 nm within the photonic band gap of the structure (i.e. the intense transmission valley between 1100 nm and 1600 nm). The defect mode is magnified in the inset of Figure 1. Instead, for the metallic $VO_2$ based microcavity the defect mode is suppressed. Noteworthy, the transmission at wavelengths longer than 1600 nm is weaker in the metallic $VO_2$ because of its infrared absorption.

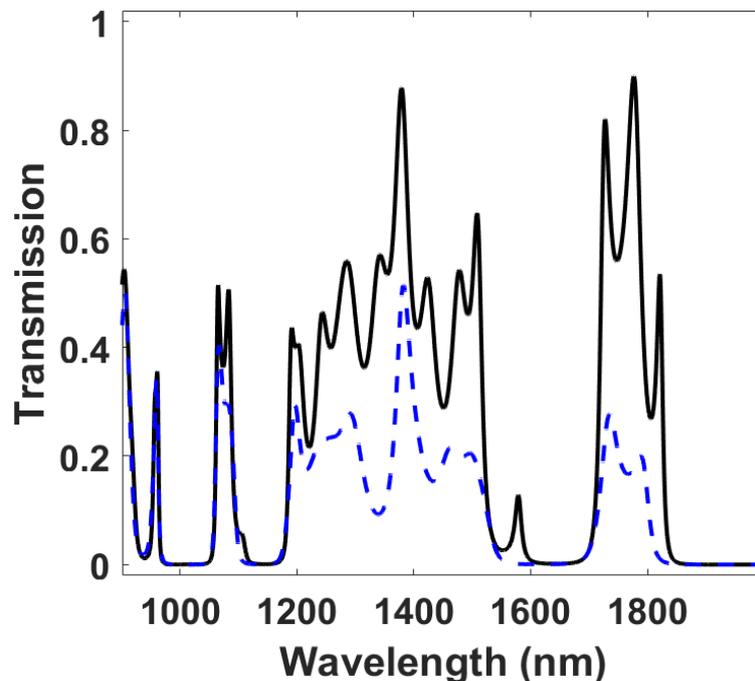

**Figure 2**. Light transmission spectra for the Thue-Morse aperiodic microcavity ABBABAABBAABABBABAABABBAABBABAAB/$VO_2$/ABBABAABBAABABBABAABABBAAB

BABAAB (A = SiO$_2$; B = ZrO$_2$) with VO$_2$ in the insulating phase (at 30 °C, black solid curve) and in the metallic phase (at 100 °C, blue dashed curve).

In Figure 2 the transmission spectrum for the aperiodic microcavity that follows the Thue-Morse sequence is depicted. VO$_2$ is sandwiched between two photonic structures with the sequence of layers ABBABAABBAABABBABAABABBAABBABAAB [23] (A = SiO$_2$; B = ZrO$_2$). The layer thicknesses are the same of the ones of the periodic structure. The black solid curve is related to VO$_2$ in the insulating phase, while the blue dashed curve is related to VO$_2$ in the metallic phase. The transmission spectra are slightly modified with the VO$_2$ from the insulating to the metallic phase, with the suppression of peaks around 1100 and 1600 nm.

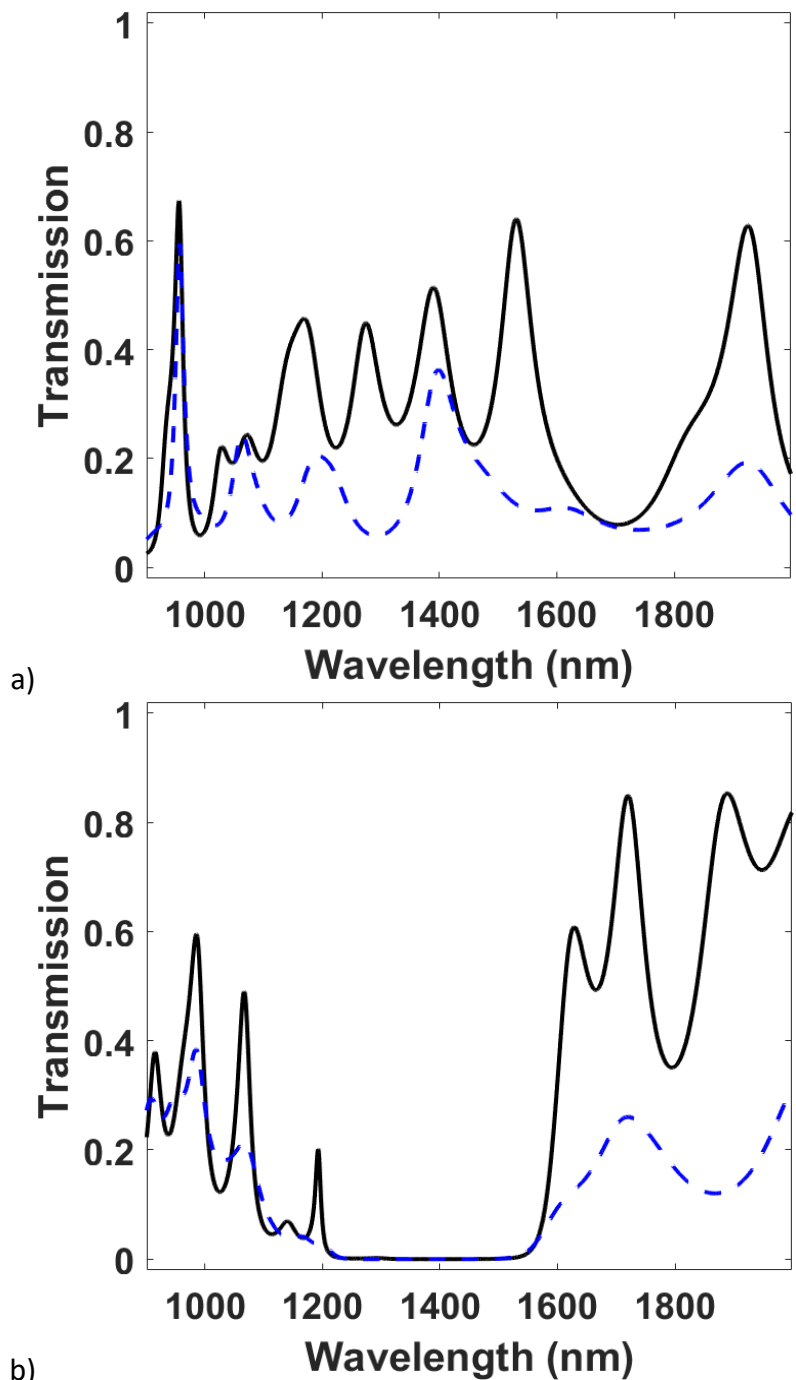

a)

b)

**Figure 3**. a) Light transmission spectra for the disordered microcavity BBABBAABBBABBBBA/VO$_2$/BABAABBBBABBBABA (A = SiO$_2$; B = ZrO$_2$) with VO$_2$ in the insulating phase (at 30 °C, black solid curve) and in the metallic phase (at 100 °C, blue dashed curve). b) Light transmission spectra for the microcavity $(SiO_2/ZrO_2)_5/VO_2/(ZrO_2/SiO_2)_5$, in which a random variation of the thicknesses is introduced, with $VO_2$ in the insulating phase (at 30 °C, black solid curve) and in the metallic phase (at 100 °C, blue dashed curve).

Also in Figure 3 the black solid curves correspond to the insulating phase of $VO_2$, while the blue dashed curves to the metallic phase of $VO_2$. In Figure 3a the transmission spectra for the disordered microcavity, in which $VO_2$ is embedded between one-dimensional random photonic structures [24,25]. The proposed structure follows the sequence BBABBAABBBABBBBA/VO$_2$/BABAABBBBABBBABA. Also in this case, the layer thicknesses are the same of the ones of the periodic structure. The transmission spectrum with $VO_2$ in the insulating phase shows eight peaks in the studied range (900 – 2000 nm). With the transition from insulator to metal the suppression of most of the transmission peaks is noticeable.

In Figure 3b is shown the transmission spectra of the microcavity $(SiO_2/ZrO_2)_5/VO_2/(ZrO_2/SiO_2)_5$, in which a random variation of the thicknesses is introduced [26,27]. For the $SiO_2$ layers the thickness is $[2 \times (110 \pm n)]$, while for the $ZrO_2$ layers the thickness is $[1.5 \times (110 \pm n)]$, where *n* is an integer random number between 0 and 20. The transmission spectrum for the microcavity with vanadium dioxide in the insulating phase shows a manifold of transmission valleys and peaks. The transition from insulator to metal suppresses several transmission peaks, as for example the narrow peak at 1200 nm.

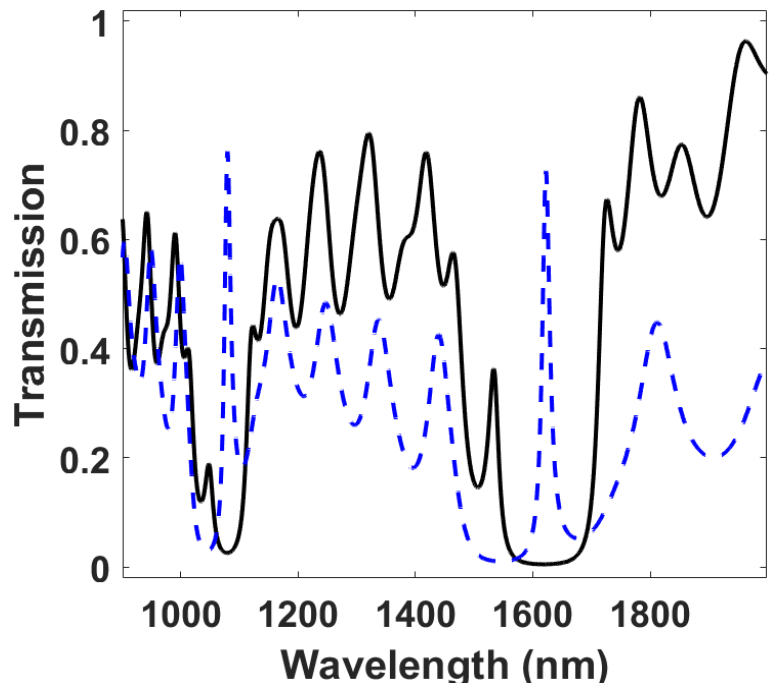

**Figure 4**. Light transmission spectra for the four material photonic crystal based microcavity $(SiO_2/Al_2O_3/Y_2O_3/ZrO_2)_6/VO_2/(ZrO_2/Y_2O_3/Al_2O_3/SiO_2)_6$ with VO2 in the insulating phase (at 30 °C, black solid curve) and in the metallic phase (at 100 °C, blue dashed curve).

In Figure 4 the transmission spectrum of the four material-based microcavity $(SiO_2/Al_2O_3/Y_2O_3/ZrO_2)_6/VO_2/(ZrO_2/Y_2O_3/Al_2O_3/SiO_2)_6$ is shown. The thickness of $SiO_2$ layers is 275.9 nm, the thickness of $Al_2O_3$ layers is 228.6 nm, the thickness of $Y_2O_3$ layers is 210.5 nm, the thickness of $ZrO_2$ layers is 190.5 nm, and the thickness of $VO_2$ layers is 45.9 nm. As shown in previous reports, four-material photonic crystals show a manifold of gaps [28]. In fact, with this structure, in the wavelength interval between 900 nm and 2000 nm two intense photonic band gaps are observable, compared to the single photonic band gap of the microcavity (Figure 1). In this case, the defect modes of the two photonic band gaps show a red shift and a remarkable intensity increase.

**Conclusion**

In this work it has been studied the light transmission of different one-dimensional photonic microcavities that embed vanadium dioxide by means of the transfer matrix method. The four types of microcavities include periodic photonic crystals, aperiodic structures, disordered structures, and four-material-based photonic crystals. The transmission of $VO_2$ from insulator to metal, achievable via a temperature increase, leads to a modulation of the transmission spectra, noticeable with shifts and suppressions of transmission peaks. The modulation of the transmission spectra of the microcavities can be exploited for smart windows and temperature controlled switches. Moreover, the photonic structure can be also used as temperature sensor since it has been studied by Currie et al. the temperature dependent refractive index dispersion of vanadium dioxide [3].

**References**

[1] F.J. Morin, Oxides Which Show a Metal-to-Insulator Transition at the Neel Temperature, Phys. Rev. Lett. 3 (1959) 34–36. https://doi.org/10.1103/PhysRevLett.3.34.

[2] R.M. Briggs, I.M. Pryce, H.A. Atwater, Compact silicon photonic waveguide modulator based on the vanadium dioxide metal-insulator phase transition, Opt. Express, OE. 18 (2010) 11192–11201. https://doi.org/10.1364/OE.18.011192.

[3] M. Currie, M.A. Mastro, V.D. Wheeler, Characterizing the tunable refractive index of vanadium dioxide, Opt. Mater. Express, OME. 7 (2017) 1697–1707. https://doi.org/10.1364/OME.7.001697.

[4] H.W. Verleur, A.S. Barker, C.N. Berglund, Optical Properties of V${\mathrm{O}}_{2}$ between 0.25 and 5 eV, Phys. Rev. 172 (1968) 788–798. https://doi.org/10.1103/PhysRev.172.788.

[5] K. Liu, S. Lee, S. Yang, O. Delaire, J. Wu, Recent progresses on physics and applications of vanadium dioxide, Materials Today. 21 (2018) 875–896. https://doi.org/10.1016/j.mattod.2018.03.029.

[6] H. Lu, S. Clark, Y. Guo, J. Robertson, The metal–insulator phase change in vanadium dioxide and its applications, Journal of Applied Physics. 129 (2021) 240902. https://doi.org/10.1063/5.0027674.

[7] S. John, Strong localization of photons in certain disordered dielectric superlattices, Phys. Rev. Lett. 58 (1987) 2486–2489. https://doi.org/10.1103/PhysRevLett.58.2486.

[8] E. Yablonovitch, Inhibited Spontaneous Emission in Solid-State Physics and Electronics, Phys. Rev. Lett. 58 (1987) 2059–2062. https://doi.org/10.1103/PhysRevLett.58.2059.

[9] J.D. Joannopoulos, ed., Photonic crystals: molding the flow of light, 2nd ed, Princeton University Press, Princeton, 2008.

[10] C. Toccafondi, L. Occhi, O. Cavalleri, A. Penco, R. Castagna, A. Bianco, C. Bertarelli, D. Comoretto, M. Canepa, Photochromic and photomechanical responses of an amorphous diarylethene-based polymer: a spectroscopic ellipsometry investigation of ultrathin films, J. Mater. Chem. C. 2 (2014) 4692–4698. https://doi.org/10.1039/C4TC00371C.

[11] P. Guo, R.D. Schaller, L.E. Ocola, B.T. Diroll, J.B. Ketterson, R.P.H. Chang, Large optical nonlinearity of ITO nanorods for sub-picosecond all-optical modulation of the full-visible spectrum, Nat Commun. 7 (2016) 12892_1-12892_10. https://doi.org/10.1038/ncomms12892.

[12]   I. Kriegel, C. Urso, D. Viola, L. De Trizio, F. Scotognella, G. Cerullo, L. Manna, Ultrafast Photodoping and Plasmon Dynamics in Fluorine–Indium Codoped Cadmium Oxide Nanocrystals for All-Optical Signal Manipulation at Optical Communication Wavelengths, J. Phys. Chem. Lett. 7 (2016) 3873–3881. https://doi.org/10.1021/acs.jpclett.6b01904.

[13]   I. Kriegel, F. Scotognella, Light-induced switching in pDTE–FICO 1D photonic structures, Optics Communications. 410 (2018) 703–706. https://doi.org/10.1016/j.optcom.2017.11.019.

[14]   Y.G. Boucher, A. Chiasera, M. Ferrari, G.C. Righini, Photoluminescence spectra of an optically pumped erbium-doped micro-cavity with SiO2/TiO2 distributed Bragg reflectors, Journal of Luminescence. 129 (2009) 1989–1993. https://doi.org/10.1016/j.jlumin.2009.04.085.

[15]   M. Born, E. Wolf, A.B. Bhatia, P.C. Clemmow, D. Gabor, A.R. Stokes, A.M. Taylor, P.A. Wayman, W.L. Wilcock, Principles of Optics: Electromagnetic Theory of Propagation, Interference and Diffraction of Light, 7th ed., Cambridge University Press, 1999. https://doi.org/10.1017/CBO9781139644181.

[16]   X. Xiao, W. Wenjun, L. Shuhong, Z. Wanquan, Z. Dong, D. Qianqian, G. Xuexi, Z. Bingyuan, Investigation of defect modes with Al2O3 and TiO2 in one-dimensional photonic crystals, Optik. 127 (2016) 135–138. https://doi.org/10.1016/j.ijleo.2015.10.005.

[17]   G.M. Paternò, L. Moscardi, S. Donini, D. Ariodanti, I. Kriegel, M. Zani, E. Parisini, F. Scotognella, G. Lanzani, Hybrid One-Dimensional Plasmonic–Photonic Crystals for Optical Detection of Bacterial Contaminants, J. Phys. Chem. Lett. 10 (2019) 4980–4986. https://doi.org/10.1021/acs.jpclett.9b01612.

[18]   I.H. Malitson, Interspecimen Comparison of the Refractive Index of Fused Silica*,†, J. Opt. Soc. Am., JOSA. 55 (1965) 1205–1209. https://doi.org/10.1364/JOSA.55.001205.

[19]   C.Z. Tan, Determination of refractive index of silica glass for infrared wavelengths by IR spectroscopy, Journal of Non-Crystalline Solids. 223 (1998) 158–163. https://doi.org/10.1016/S0022-3093(97)00438-9.

[20]   I.H. Malitson, Refraction and Dispersion of Synthetic Sapphire, J. Opt. Soc. Am., JOSA. 52 (1962) 1377–1379. https://doi.org/10.1364/JOSA.52.001377.

[21]   Y. Nigara, Measurement of the Optical Constants of Yttrium Oxide, Jpn. J. Appl. Phys. 7 (1968) 404. https://doi.org/10.1143/JJAP.7.404.

[22]   D.L. Wood, K. Nassau, Refractive index of cubic zirconia stabilized with yttria, Appl. Opt., AO. 21 (1982) 2978–2981. https://doi.org/10.1364/AO.21.002978.

[23]   W. Steurer, D. Sutter-Widmer, Photonic and phononic quasicrystals, J. Phys. D: Appl. Phys. 40 (2007) R229–R247. https://doi.org/10.1088/0022-3727/40/13/R01.

[24]   D.S. Wiersma, Disordered photonics, Nature Photonics. 7 (2013) 188–196. https://doi.org/10.1038/nphoton.2013.29.

[25]   D.S. Wiersma, R. Sapienza, S. Mujumdar, M. Colocci, M. Ghulinyan, L. Pavesi, Optics of nanostructured dielectrics, J. Opt. A: Pure Appl. Opt. 7 (2005) S190–S197. https://doi.org/10.1088/1464-4258/7/2/025.

[26]   J. Faist, J. -D. Ganière, Ph. Buffat, S. Sampson, F. -K. Reinhart, Characterization of GaAs/(GaAs)n(AlAs)m surface-emitting laser structures through reflectivity and high-resolution electron microscopy measurements, Journal of Applied Physics. 66 (1989) 1023–1032. https://doi.org/10.1063/1.343488.

[27]   A. Chiasera, F. Scotognella, L. Criante, S. Varas, G.D. Valle, R. Ramponi, M. Ferrari, Disorder in Photonic Structures Induced by Random Layer Thickness, Sci Adv Mater. 7 (2015) 1207–1212. https://doi.org/10.1166/sam.2015.2249.

[28]   I. Kriegel, F. Scotognella, Band gap splitting and average transmission lowering in ordered and disordered one-dimensional photonic structures composed by more than two materials

with the same optical thickness, Optics Communications. 338 (2015) 523–527. https://doi.org/10.1016/j.optcom.2014.10.045.